\documentclass[preprint,superscriptaddress,notitlepage,endfloat]{revtex4-2}
\usepackage{graphicx,bm,mathtools,color}
\usepackage[pdfstartview=FitH, CJKbookmarks=true, bookmarksnumbered=true, bookmarksopen=true, colorlinks, pdfborder=001, linkcolor=blue, anchorcolor=blue, citecolor=blue]{hyperref}
\usepackage{ulem}
\usepackage{verbatim}

\newcommand{\del}[1]{\iffalse #1 \fi}
\newcommand{\be}{\begin{equation}}
	\newcommand{\ee}{\end{equation}}

\newcommand{\CCG}{Ce$_{5}$CoGe$_{2}$}
\newcommand{\hide}[1]{}

\begin{document}
	
        \renewcommand{\abstractname}{} 
		\title{Pressure-induced superconductivity beyond magnetic quantum criticality in a Kondo ferromagnet}

	\author{Yanan Zhang}
	\thanks{These authors contributed equally to this work.}
\affiliation  {New Cornerstone Science Laboratory, Center for Correlated Matter and School of Physics, Zhejiang University, Hangzhou 310058, China}
	\author{Yongjun Zhang} 
	\thanks{These authors contributed equally to this work.}
	\affiliation  {Hubei Key Laboratory of Photoelectric Materials and Devices, School of Materials Science and Engineering, Hubei Normal University, Huangshi 435002, China}
	
		\author{Jiawen Zhang}
        \thanks{These authors contributed equally to this work.}
	\affiliation  {New Cornerstone Science Laboratory, Center for Correlated Matter and School of Physics, Zhejiang University, Hangzhou 310058, China}

		\author{Kaixin Ye}
		\affiliation  {New Cornerstone Science Laboratory, Center for Correlated Matter and School of Physics, Zhejiang University, Hangzhou 310058, China}

		\author{Dajun Su} 
		\affiliation  {New Cornerstone Science Laboratory, Center for Correlated Matter and School of Physics, Zhejiang University, Hangzhou 310058, China}

    	\author{Yanen Huang} 
		\affiliation  {New Cornerstone Science Laboratory, Center for Correlated Matter and School of Physics, Zhejiang University, Hangzhou 310058, China}
			
		\author{Zhaoyang Shan}
		\affiliation  {New Cornerstone Science Laboratory, Center for Correlated Matter and School of Physics, Zhejiang University, Hangzhou 310058, China}

	\author{Jiyuan Li} 
    \affiliation  {Hubei Key Laboratory of Photoelectric Materials and Devices, School of Materials Science and Engineering, Hubei Normal University, Huangshi 435002, China}
				
		\author{Rui Li}
		\affiliation  {New Cornerstone Science Laboratory, Center for Correlated Matter and School of Physics, Zhejiang University, Hangzhou 310058, China}
		
		\author{Ye Chen}
	\affiliation  {New Cornerstone Science Laboratory, Center for Correlated Matter and School of Physics, Zhejiang University, Hangzhou 310058, China}
		
			\author{Xin Lu}
	\affiliation  {New Cornerstone Science Laboratory, Center for Correlated Matter and School of Physics, Zhejiang University, Hangzhou 310058, China}
		
	\author{Lin Jiao}
	\affiliation  {New Cornerstone Science Laboratory, Center for Correlated Matter and School of Physics, Zhejiang University, Hangzhou 310058, China}
		
	\author{Yu Liu}
    \email[Corresponding author: ]{liuyuccm@zju.edu.cn}
	\affiliation  {New Cornerstone Science Laboratory, Center for Correlated Matter and School of Physics, Zhejiang University, Hangzhou 310058, China}

		\author{Michael Smidman}
        \email[Corresponding author: ]{msmidman@zju.edu.cn}
		\affiliation  {New Cornerstone Science Laboratory, Center for Correlated Matter and School of Physics, Zhejiang University, Hangzhou 310058, China}	
		 \author{Frank Steglich}
	\affiliation  {New Cornerstone Science Laboratory, Center for Correlated Matter and School of Physics, Zhejiang University, Hangzhou 310058, China}
		 \affiliation  {Max Planck Institute for Chemical Physics of Solids (MPI CPfS), 01187 Dresden, Germany}
		\author{Huiqiu Yuan}
		\email[Corresponding author: ]{hqyuan@zju.edu.cn}
	\affiliation  {New Cornerstone Science Laboratory, Center for Correlated Matter and School of Physics, Zhejiang University, Hangzhou 310058, China}
		\affiliation  {Institute of Fundamental and Transdisciplinary Research, Zhejiang University, Hangzhou 310058, China}
		\affiliation  {Institute for Advanced Study in Physics, Zhejiang University, Hangzhou 310058, China}
	\affiliation  {State Key Laboratory of Silicon and Advanced Semiconductor Materials, Zhejiang University, Hangzhou 310058, China}
		\affiliation  {Collaborative Innovation Center of Advanced Microstructures, Nanjing 210093, China}

	\date{\today}

\begin{abstract} 
\textbf{Quantum phase transitions are an established setting for emergent phenomena driven by strong electronic correlations, including strange metals and unconventional superconductivity. These have been explored extensively in Kondo lattice materials tuned to an antiferromagnetic quantum critical point (QCP), but superconductivity emerging near ferromagnetic quantum criticality is not yet observed, and the conditions under which it occurs in proximity to ferromagnetism are undetermined. Here, we report a new setting for superconductivity in the ferromagnetic Kondo-lattice material Ce$_5$CoGe$_2$, where there is a ferromagnetic ground state at ambient pressure, which evolves to antiferromagnetism under applied pressures. The antiferromagnetic transition is suppressed to a zero-temperature QCP, which is accompanied by strange-metal behavior. Superconductivity does not occur at the QCP, but instead appears at pressures beyond the magnetic instability. These findings suggest that Ce$_5$CoGe$_2$ represents a distinct class of correlated materials exhibiting a unique scenario for the emergence of superconductivity, likely associated with unconventional pairing mechanisms beyond spin-fluctuations. 
}

\end{abstract}

	\maketitle
\begin{figure}[h]
	\begin{center}
		\includegraphics[width=0.8\columnwidth]{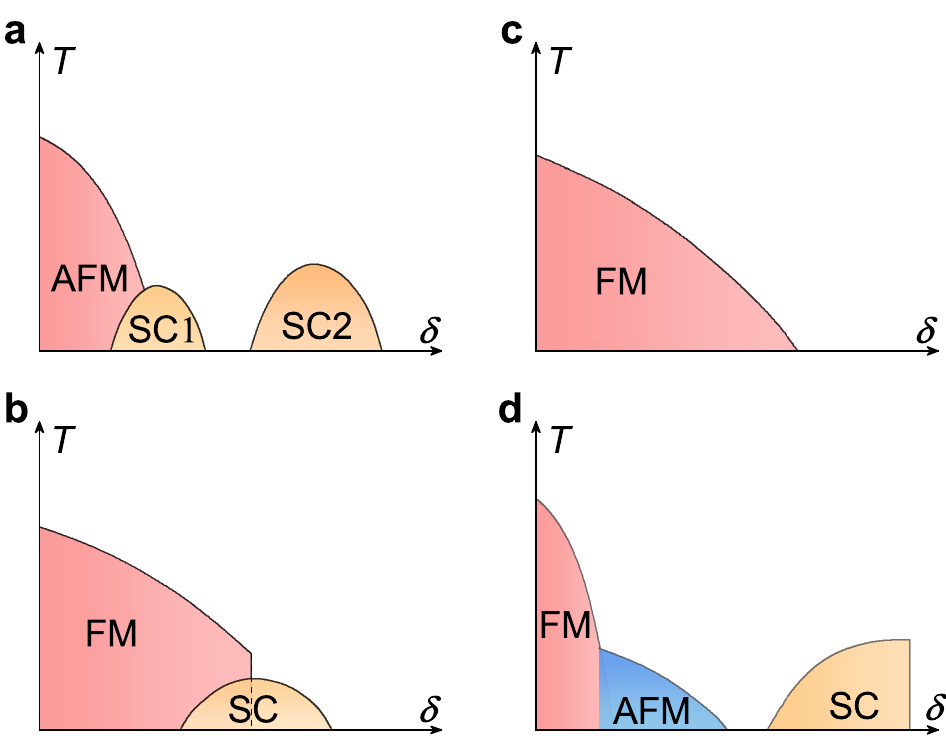}
	\end{center}
	\caption{\textbf{Schematic phase diagrams of superconductivity and magnetism in different quantum materials.}
		\textbf{a,} Representative phase diagram of CeCu$_2$Si$_2$, where the first superconducting dome (SC1) emerges near the antiferromagnetic (AFM) quantum critical point (QCP). As the tuning parameter $\delta$ increases, a second superconducting dome (SC2) appears, likely driven by valence fluctuations \cite{Yuan2003,Yuan2006}. 
		\textbf{b,} Schematic phase diagram of uranium-based ferromagnetic superconductors, where superconductivity (SC) coexists with ferromagnetic (FM) order. In these systems, the FM transition is typically first-order, avoiding a QCP \cite{Saxena2000}.
		\textbf{c,} Schematic phase diagram of a continuous FM QCP, at which superconductivity is not yet observed \cite{steppke2013,shen2020N}.
	    \textbf{d,} Schematic phase diagram of Ce$_5$CoGe$_2$ under pressure. With increasing pressure, FM order first gives way to AFM order, which is subsequently continuously suppressed to zero temperature at an AFM QCP. Superconductivity emerges at higher pressures beyond the AFM QCP.
	}
	
	\label{Fig1}
\end{figure}

The close proximity of unconventional superconductivity to magnetism has been revealed in a variety of materials settings, including heavy-fermion superconductors \cite{Pfleiderer-RMP}, high-temperature cuprate \cite{lee2006RMP} and iron-based superconductors \cite{Dai-rmp}, and organic superconductors \cite{Lang2004}, which has been interpreted in the framework of  the superconductivity being driven by spin fluctuations \cite{rmp2012,stewart2017}. This paradigm has been extensively investigated in heavy-fermion antiferromagnets \cite{Mathur1998,Schrder2000,2001StewartRMP,2007LohneysenRMP,Gegenwart2008,Michael_RMP}, where the relative strengths of the competing Ruderman–Kittel–Kasuya–Yosida (RKKY) and Kondo interactions are tuned by experimentally applying non-thermal control parameters, and unconventional superconductivity often emerges upon the suppression of antiferromagnetic order at a QCP. Moreover, in such systems superconductivity may also arise driven by other degrees of freedom, such  as quadrupolar \cite{Sakai2012} or valence fluctuations \cite{Yuan2003,Bauer2011,Matsumoto2011}, where the latter has been proposed to explain the second superconducting state of pressurized CeCu$_2$Si$_2$, which is well separated from the AFM QCP \cite{Yuan2003,Yuan2006,valence_CCS_2004}, as  illustrated in Fig.~\ref{Fig1}a.

In contrast, the evidence for superconductivity arising from ferromagnetic (FM) quantum criticality is more limited \cite{BrandoRMP2016}. Although several uranium-based compounds exhibit the coexistence of superconductivity and FM order (Fig.~\ref{Fig1}b), in these cases the FM transition terminates abruptly under pressure at a first-order transition, avoiding  a  QCP \cite{Saxena2000,Aoki2001,Slooten2009}. On the other hand, while pressure and doping have been reported to induce continuous FM QCPs in CeRh$_6$Ge$_4$ \cite{shen2020N,Wang2021} and YbNi$_4$(P$_{1-x}$As$_x$)$_2$ \cite{steppke2013}, respectively (Fig.~\ref{Fig1}c), no superconductivity has yet been found in such materials. Another scenario whereby FM quantum criticality is avoided is by a change of ground state from FM order, to  spin-density wave (SDW) or other forms of AFM ground state upon tuning with pressure or doping. Typical examples include the Kondo lattice materials CeRu$_2$Ge$_2$ \cite{sullow1999}, CeAgSb$_2$ \cite{Sidorov2003} and CeRuPO \cite{2015CeRuPO} as well as transition metal compounds NbFe$_2$ \cite{2008Brando} and LaCrGe$_3$ \cite{2016LaCrGe}, in which  superconductivity is also not yet observed.

Here, we report a new superconductor Ce$_5$CoGe$_2$, which corresponds to a novel setting for superconductivity in proximity to magnetism. At ambient pressure, Ce$_5$CoGe$_2$ is a Kondo lattice compound showing the coexistence of ferromagnetism and cluster glass behavior below the Curie temperature ($T_{\mathrm{\mathrm{\mathrm{C}}}}$) of 10.9~K (Extended Data Fig. 1)~\cite{su2024}. The resulting  temperature-pressure phase diagram is shown schematically  in Fig.~\ref{Fig1}d, where under pressure the FM order first gives way to AFM order. With further increasing pressure, the AFM order is continuously suppressed and vanishes at an AFM QCP. Interestingly, superconductivity does not emerge immediately upon the suppression of antiferromagnetic order, but appears at higher pressures, being separated from the AFM instability.

\begin{figure}[h]
	\begin{center}
		\includegraphics[width=1.0\columnwidth]{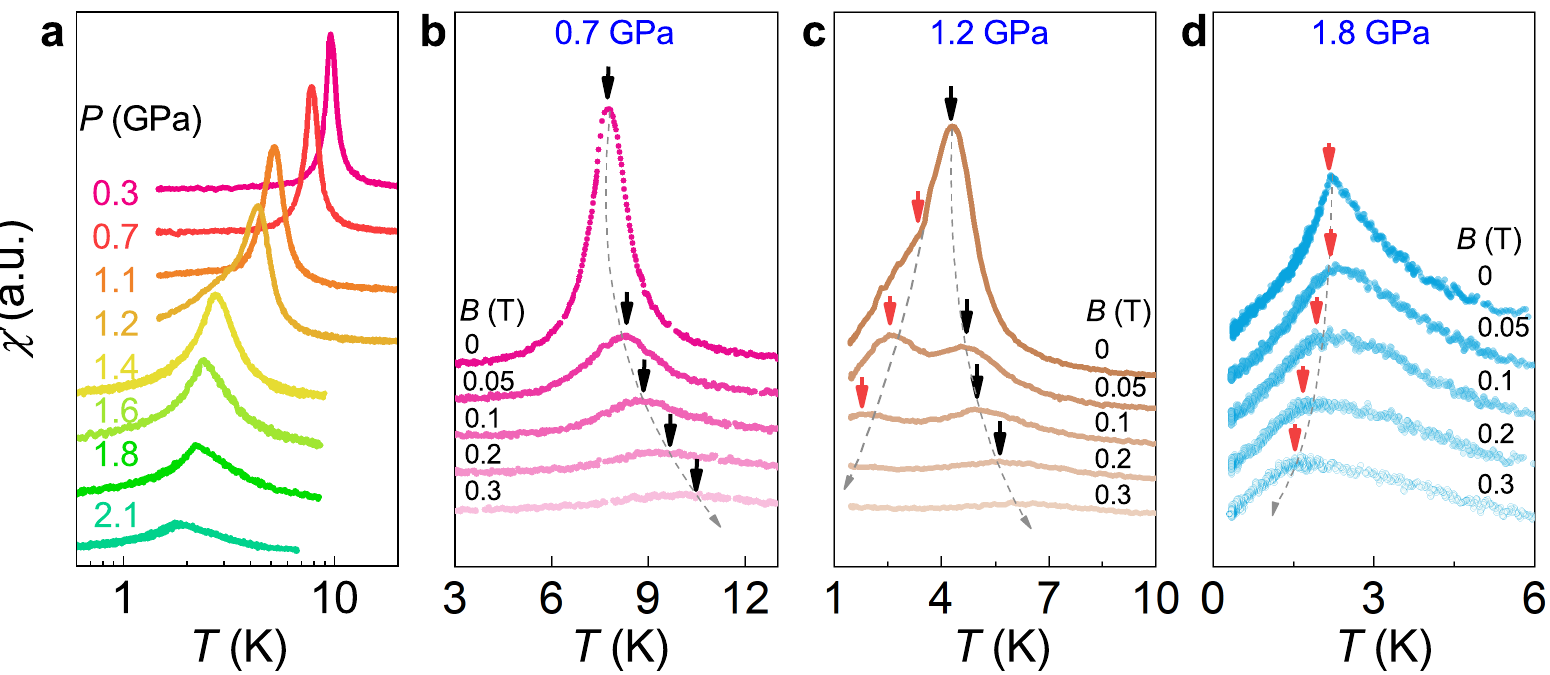}
	\end{center}
	\caption{\textbf{Pressure-induced FM to AFM transition in \CCG.} \textbf{a,} Temperature dependence of the real part of the ac susceptibility $\chi^{\prime}(T)$ of {Ce$_{5}$CoGe$_{2}$} under various pressures from 0.3 to 2.1 GPa. $\chi^{\prime}(T)$ of {Ce$_{5}$CoGe$_{2}$} is displayed for \textbf{b,} 0.7 GPa, \textbf{c,} 1.2 GPa, and \textbf{d,} 1.8 GPa, under various applied magnetic fields up to 0.3 T. The black and red arrows indicate FM and AFM transitions, respectively. Note that the curves are vertically shifted for clarity.}

	\label{Fig2}
\end{figure}

In order to track the evolution of the magnetic ground state of \CCG\ with pressure, the temperature dependence of the ac magnetic susceptibility was measured under pressure, and the real part $\chi'(T)$ is displayed in Fig.~\ref{Fig2}a under pressures up to 2.1 GPa. With increasing pressure, the peak in $\chi'(T)$ corresponding to the magnetic transition shifts from 9.7 K at ambient pressure to 1.7 K at 2.1 GPa, demonstrating a pressure induced suppression of magnetic order. Meanwhile, measurements in applied magnetic fields (Figs.~\ref{Fig2}b-d) reveal distinct changes in the magnetic ground state under pressure. At  lower pressures such as 0.7 GPa, the transition moves to higher temperatures with increasing field, in line with a ferromagnetic transition, while at 1.8 GPa, the transition shifts to lower temperatures, which is characteristic of AFM order.  At the intermediate pressure of 1.2 GPa, two peaks are observed with different field dependences, indicating the coexistence of FM ($T_{\mathrm{\mathrm{\mathrm{C}}}} \approx 4.3$~K) and AFM ($T_{\mathrm{\mathrm{N}}} \approx 3.5$~K) transitions at this pressure. Furthermore, the frequency dependence of $\chi'(T)$ indicates that the spin-cluster behavior present at ambient pressure \cite{su2024} vanishes within the AFM phase (Extended Data Fig. 2). These results suggest that upon applying pressure to \CCG\, there is a change from an FM to AFM magnetic ground state. The nature of the coexistence of FM and AFM at 1.2 GPa is not yet determined; it could be that the FM-AFM boundary is first order and there are coexistent macroscopic FM and AFM domains in its vicinity, or it could be there is a more microscopic coexistence, for example if the ordering on the 4 inequivalent Ce sites evolves differently with pressure.

\begin{figure}[h]
	\begin{center}
		\includegraphics[width=1.0\columnwidth]{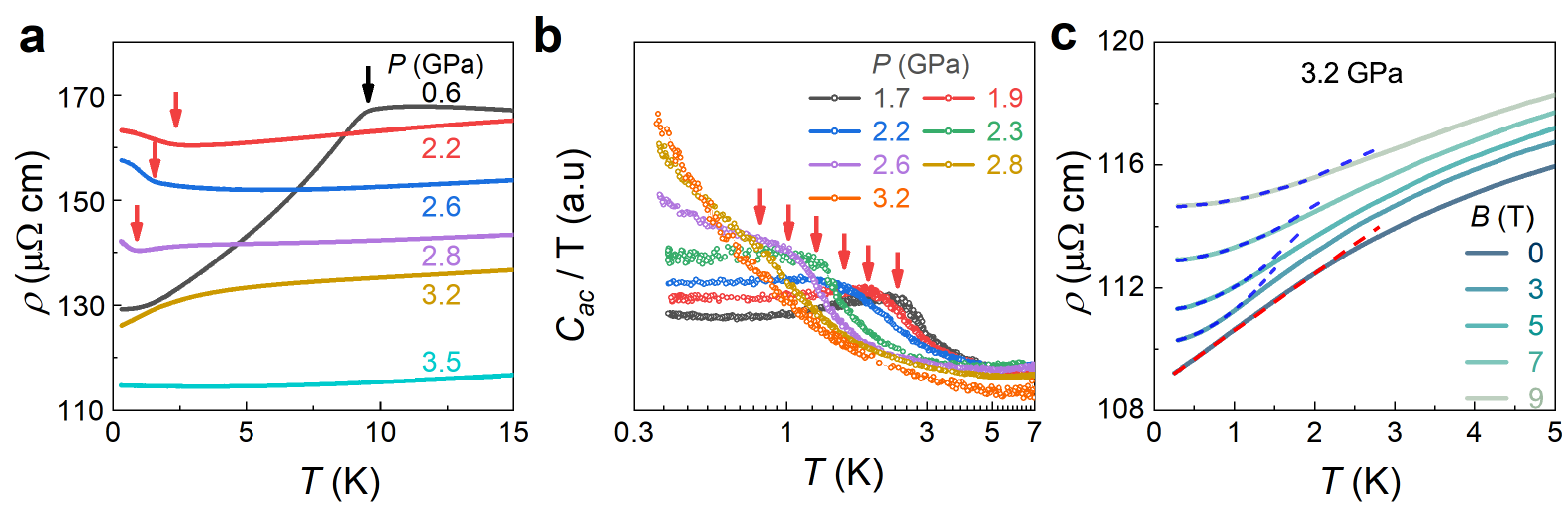}
	\end{center}
 \caption{\textbf{Quantum critical behavior in \CCG.} \textbf{a,} Low temperature dependence of resistivity $\rho(T)$ of \CCG\ measured at pressures between 0.6 and 3.5 GPa. The red and black arrows indicate AFM and FM transitions, respectively. \textbf{b,} Temperature dependence of the ac heat capacity coefficient $C_{\rm ac}(T)/T$ of \CCG\ measured at pressures between 1.7 and 3.2~GPa, where the red arrows indicate AFM transitions. \textbf{c,} $\rho(T)$ under various applied magnetic fields at 3.2 GPa. The red dashed line marks the $T$-linear resistivity, corresponding to strange-metal behavior. The blue dashed lines show fits to a $T^{\rm 2}$ dependence, corresponding to Fermi liquid behavior.
 }
	
	\label{Fig3}
\end{figure}

The electrical resistivity $\rho(T)$ of \CCG\  is displayed in Fig.~\ref{Fig3}a and Extended Data Fig. 3.
\CCG\ exhibits a  more metallic behavior as the pressure is increased from 0.6 GPa to 5~GPa (Extended Data Fig. 3a).   The magnetic transition is also detected in the low temperature $\rho(T)$ (Fig.~\ref{Fig3}a), where the FM transition appears as a kink below which $\rho(T)$ decreases rapidly, and $T_{\rm C}$ can be defined as the position of the maximum in the derivative $d\rho(T)/dT$ (Extended Data Fig. 4). On the other hand, there is an upturn in  $\rho(T)$ upon cooling below the AFM transition, and the Neel temperature $T_{\rm N}$ corresponds to the minimum in $d\rho(T)/dT$ (Extended Data Fig. 4), which could signal the opening of a gap  \cite{Luo2015}. Additional resistivity measurements in a piston–cylinder cell confirm the reproducibility of the resistivity upturn (Extended Data Fig. 5). The upturn in $\rho(T)$ moves to lower temperatures with increasing pressure, which together with the  $\chi'(T)$ (Fig.~\ref{Fig2}a) and ac heat capacity results (Fig.~\ref{Fig3}b and Extended Data Fig. 6), demonstrates that the AFM transition is continuously suppressed to zero temperature, reaching a QCP at a critical pressure of $p_{\rm c}\approx3.2$~GPa.  At $p_{\rm c}$, $C_{\rm ac}(T)/T$ diverges, while the resistivity is linear in temperature from 2 K down to at least 250 mK (Fig.~\ref{Fig3}c), indicating strange-metal behavior at the QCP. When magnetic fields are applied along the $a$ axis, there is a change from a $\sim T$ to $\sim T^2$ dependence of $\rho(T)$, suggesting the restoration of Fermi liquid behavior in applied fields (Extended Data Fig. 7).

\begin{figure}[h]
		\begin{center}
			\includegraphics[width=0.8\columnwidth]{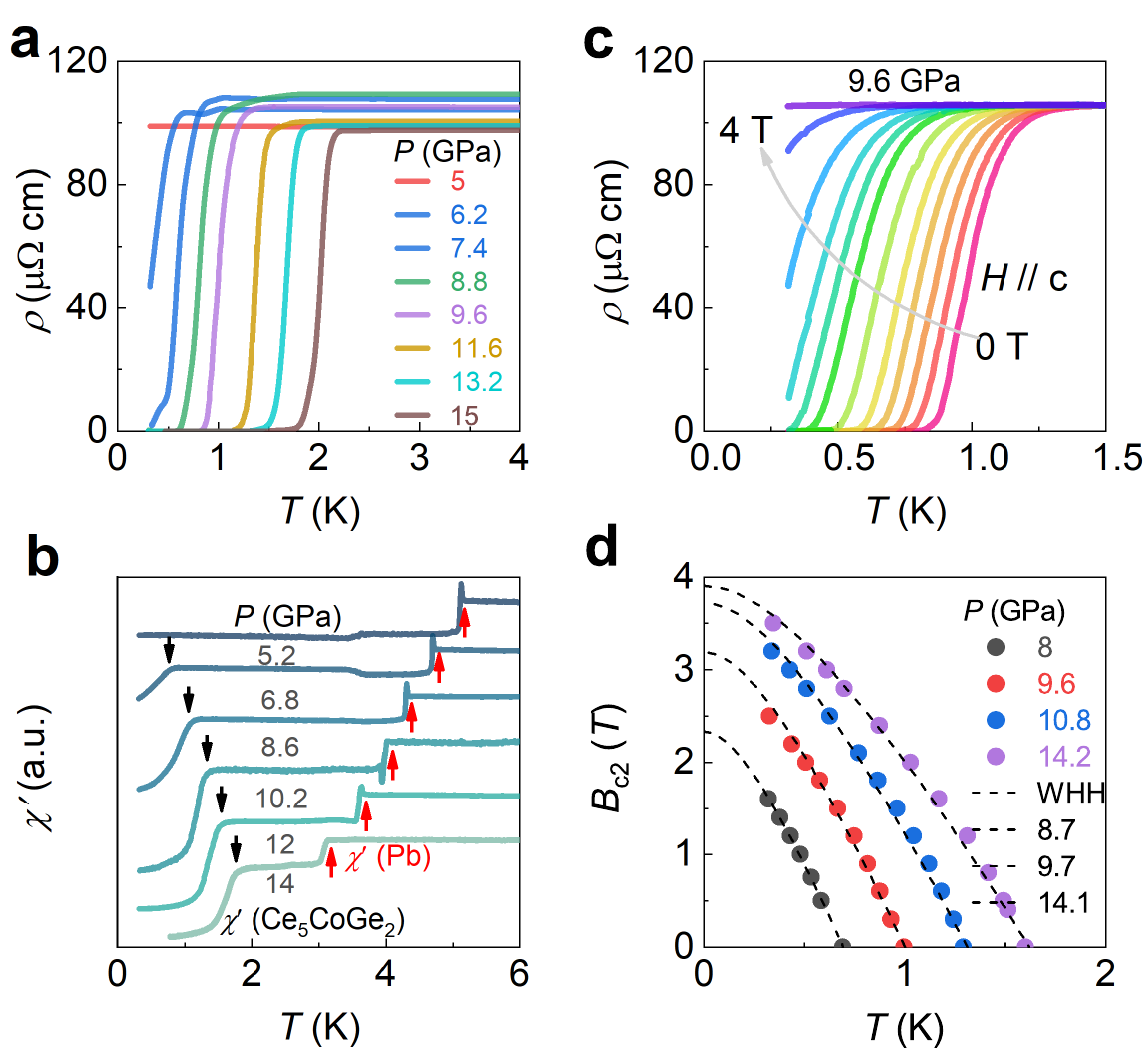}
		\end{center}
	\caption{\textbf{Superconductivity of \CCG.}  \textbf{a,} Low temperature $\rho$($T$) of \CCG\ between 5 and 15~GPa.  \textbf{b,} $\chi^{\prime}(T)$ of \CCG\ at different pressures, where superconducting transitions corresponding to the sample and a Pb piece for reference are marked by black and red arrows, respectively. \textbf{c,} $\rho(T)$ under various applied magnetic fields at 9.6 GPa. \textbf{d,} Temperature dependence of the upper critical fields ($B_{\mathrm{c2}}$) at different pressures, with black dashed lines representing the WHH model fits.}

		\label{Fig4}
	\end{figure}

Figure \ref{Fig4}a and Extended Data Fig. 3b display $\rho(T)$ at higher pressures above $p_c$ up to 15 GPa. At 6.2 GPa, there is a sharp drop in $\rho(T)$ below 0.5~K, indicating the onset of superconductivity. At higher pressures, the transition temperature $T_{\rm sc}$ (determined from where $\rho(T)$ drops by 50\%) moves to higher temperatures, reaching 2 K at 15 GPa, and zero-resistance is realized at lower temperatures. To corroborate the occurrence of superconductivity, ac magnetic susceptibility measurements were performed under pressure (Fig.~\ref{Fig4}b), with a small piece of Pb in the pressure cell as a reference. The comparable jump in $\chi'(T)$ between Ce$_5$CoGe$_2$ and Pb provides clear evidence for a full shielding fraction in Ce$_5$CoGe$_2$ below the  superconducting transition. Meanwhile, the normal state resistivity remains metallic across the whole pressure range, but there is a slight upturn at low temperatures (Extended Data Fig. 8), reminiscent of CeCu$_2$Si$_2$ under pressure \cite{Yuan2003PhD}.

\begin{figure}[h]
		\begin{center}
			\includegraphics[width=0.8\columnwidth]{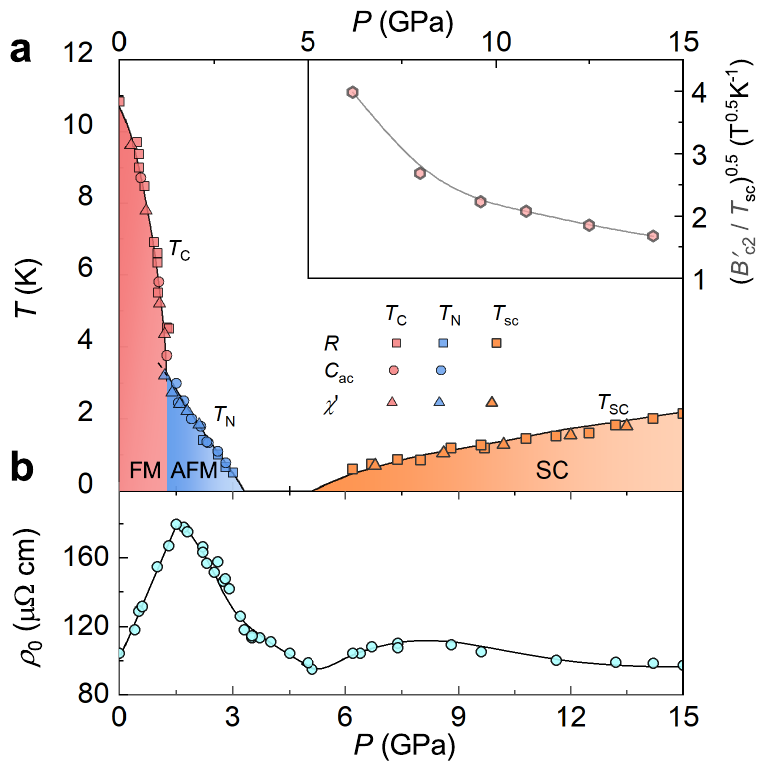}
		\end{center}
\caption{\textbf{Phase diagram of \CCG\ under pressure.} 
	\textbf{a,} Temperature–pressure phase diagram of \CCG\ based on resistivity, ac heat capacity, and ac susceptibility measurements. The pink, blue, and orange symbols represent $T_{\rm C}$, $T_{\rm N}$, and $T_{\rm sc}$, respectively. The shaded regions correspond to the different labeled phases. The inset shows the pressure dependence of $(B'_{\mathrm{c2}}/T_{\rm sc})^{0.5}$, which qualitatively describes the evolution of the effective carrier mass.
	\textbf{b,} Pressure dependence of the low-temperature resistivity $\rho_{0}$, where $\rho_0$ is defined as the resistivity at 0.3 K. For pressures at which superconductivity occurs, $\rho_0$ corresponds to the resistivity of the normal state just above the superconducting transition.
}

		\label{Fig5}
	\end{figure}

Figure~\ref{Fig4}c displays $\rho(T)$ at 9.6 GPa under different applied magnetic fields (see also Extended Data Fig. 9 for other pressure points), while the derived upper critical fields versus temperature $B_{\mathrm{\mathrm{c2}}}(T)$ are shown in Fig.~\ref{Fig4}d, which are well fitted using the Werthamer-Helfand-Hohenberg (WHH) model \cite{WHH}. At all pressures the upper critical field exceeds the weak coupling Pauli limit ($B_{\mathrm{P}}$= 1.86 $T_{\rm sc}$). For example, at 8 GPa $B_{\mathrm{P}} = 1.3~\mathrm{T}$, but the superconducting transition is still observed in applied fields above this, and an extrapolated zero-temperature value  $B_{\mathrm{c2}}(0) = 2.3~\mathrm{T}$ is obtained. The inset of Fig.~\ref{Fig5}a displays the pressure dependence of  $(B'_{\mathrm{c2}}/T_{\rm sc})^{0.5}$, where $B^{'}_{\mathrm{c2}}$ = $-$(d$B_{\mathrm{c2}}$/d$T)_{T=T_{\rm sc}}$ is the initial slope of the upper critical field. Within the free-electron approximation, this ratio is proportional to the relative enhancement of the effective charge-carrier mass over the free electron value, $m^*/m_0 \sim (B'_{\mathrm{c2}}/T_{\rm sc})^{0.5}$ \cite{knebel2008,Park2008,Squire2024}. The value of $(B'_{\mathrm{c2}}/T_{\rm sc})^{0.5}$, which is plotted versus $T_{\rm sc}$ for various heavy-fermion and other unconventional superconductors in Extended Data Fig.~10,  is comparable to canonical heavy-fermion compounds such as CeIn$_3$ \cite{Knebel2001} and CeCoIn$_5$ \cite{Tayama2002}. 
Upon further increasing the pressure,  $(B'_{\mathrm{c2}}/T_{\rm sc})^{0.5}$ and hence  $m^*$ decreases, reaching a value close to that of the intermediate valence superconductor PuCoIn$_5$ \cite{Bauer2011}, as well as CeCu$_2$Si$_2$ in the high-pressure superconducting phase (Extended Data Fig. 10) \cite{Vargoz1998}. In addition, although $m^*/m_0$ is reduced by approximately a factor of two at $14.2~\mathrm{GPa}$, it remains strongly enhanced compared with conventional BCS superconductors: $m^*/m_0$ in \CCG\ is still about $13.6$ and $5.6$ times larger than in MgB$_2$ and Nb$_3$Sn \cite{Askerzade2002, Nbsn}, respectively (Extended Data Fig. 10). This  comparison indicates that electronic correlations in \CCG\ are stronger than in typical conventional BCS superconductors.

The resulting temperature-pressure phase diagram of Ce$_5$CoGe$_2$ is shown in Fig.~\ref{Fig5}. The ground state evolves from FM order at ambient pressure, which changes to an AFM state above 1.2 GPa, and this antiferromagnetism is suppressed to a QCP at around $p_{\mathrm{c}}\approx3.2$~GPa. Above the QCP at 6.2 GPa, superconductivity emerges and $T_{\rm sc}$ increases monotonically with pressure up to at least 15 GPa. The pressure evolution of the low temperature resistivity $\rho_{0}$ is shown in Fig.~\ref{Fig5}b, where there is a pronounced peak corresponding to the FM–AFM transition. Moreover, a broad hump in $\rho_0$ appears at elevated pressures after the emergence of superconductivity.

Overall these results demonstrate that Ce$_5$CoGe$_2$ manifests a new scenario for the interplay of magnetism and superconductivity, where the superconductivity emerges in proximity to a QCP associated with the suppression of AFM order, that in turn replaces the ambient pressure FM phase under pressure. While such a change of magnetic ground state occurs in several metallic ferromagnets \cite{BrandoRMP2016}, it is not typically associated with superconductivity. Moreover, unlike canonical antiferromagnetic heavy fermion superconductors \cite{Mathur1998,park2006N,Schuberth2016,Squire2024}, the superconductivity  does not appear to directly stem from a magnetic quantum critical point, but instead the superconducting dome is separate from the ordered phase (Fig.~\ref{Fig1}d), being similar to the case of $\beta$-YbAlB$_4$ \cite{Matsumoto2011,Tomita_YbAlB_PRESSURE},  suggesting that the superconductivity is not primarily driven by spin fluctuations.

Although the superconducting pairing state of Ce$_5$CoGe$_2$ is to be explored in future studies, the Coulomb repulsion associated with such a strongly correlated underlying electronic state, as evidenced by the enhanced $m^*$, is disfavourable to conventional $s$-wave superconductivity with onsite pairing. These considerations lead to the prospect that the superconductivity is driven by an alternative instability such as valence fluctuations, which are proposed to underly the high pressure superconducting dome of CeCu$_2$Si$_2$  \cite{Yuan2003,Yuan2006,valence_CCS_2004}. An indication of the possible role of valence fluctuations in Ce$_5$CoGe$_2$ is the rapid reduction of the electronic effective mass upon increasing pressure (inset of Fig.~\ref{Fig5}a), which is characteristic of the crossover from a heavy fermion  to a intermediate valence state \cite{Yuan2006}. In addition, valence fluctuations may enhance $\rho_{0}$, as seen in pressurized CeCu$_2$(Si$_{1-x}$Ge$_x$)$_2$ ~\cite{valence_CCS_2004}, and an increase of $\rho_{0}$ is also observed above 5 GPa in Ce$_5$CoGe$_2$, concomitant with the appearance of superconductivity. Moreover, a weak upturn of the resistivity with decreasing temperature corresponding to a -log$T$ dependence  is observed in the high-pressure superconducting regimes of both Ce$_5$CoGe$_2$ and CeCu$_2$Si$_2$ \cite{Yuan2003PhD}, which  is robust against magnetic fields (Extended Data Fig. 8). Although the origin is not determined, it may be attributed to  unscreened magnetic moments in the mixed valence regime. We note that in candidate valence-fluctuation driven superconductors $T_{\rm sc}$(P) can increase over a broad pressure range without an observed maximum. For example, it has been suggested that the superconductivity of PuCoGa$_5$ with a record $T_{\rm sc}$ of 18.5~K is mediated by valence fluctuations \cite{Bauer2011}, and like \CCG, the superconducting phase of PuCoGa$_5$ is robust, with $T_{\rm sc}$ changing only slightly over a wide pressure range \cite{Griveau2004}.  It will therefore be important to extend the phase diagram to higher pressures in future measurements to establish the full evolution of the superconducting phase of \CCG.

While the upper critical field exceeding the weak-coupling Pauli limit in a centrosymmetric superconductor could hint at a non-singlet pairing state, it should be noted that other effects, namely strong coupling and an effective Land\'e factor that diverges significantly from the free electron value, may increase the paramagnetic limiting field. Therefore, it is of particular importance to both characterize the nature of the superconducting order parameter and pairing state, and also to reveal the dynamics associated with degrees of freedom such as spin and valence fluctuations that may drive superconductivity. Furthermore, it is necessary to understand the nature of the quantum criticality of Ce$_5$CoGe$_2$, where strange-metal behavior is only observed in a narrow pressure range close to $p_{\rm c}$. More broadly, such a unique scenario for the interplay of ferromagnetism, antiferromagnetism and superconductivity, together with strange metal quantum criticality, positions Ce$_5$CoGe$_2$ as a promising materials platform for exploring novel types of unconventional superconductivity in proximity to magnetic instabilities.

		\clearpage
		
	\noindent \textbf{Methods} \\
		\noindent Single crystals of Ce$_5$CoGe$_2$ were grown using a self-flux method. Cerium ingot (Alfa Aesar, 99.9\%), cobalt slug (Alfa Aesar, 99.95\%), and germanium granules (PrMat, 99.9999\%) were first arc-melted in a molar ratio of 9:3:1 under a titanium-gettered argon atmosphere. The obtained ingot was then placed in a tantalum crucible and sealed in an evacuated quartz tube. The tube was heated to 1150 °C, held at this temperature for 24 hours, and then slowly cooled to 550 °C.

		Measurements under applied pressures up to 2.3 GPa were performed using a piston-cylinder-type pressure cell with Daphne 7373 as the pressure-transmitting medium. The applied pressure was determined from the shift in $T_{\rm sc}$ of a high-quality Pb single crystal~\cite{Eiling1981}. Measurements at pressures up to 15 GPa were carried out in a diamond anvil cell (DAC), where Daphne 7373 was used as the pressure-transmitting medium. The DAC was loaded together with several small ruby balls for pressure determination at room temperature using the ruby fluorescence method \cite{mao1986rubycalibration}. Electrical resistance measurements in both the piston-cylinder and diamond anvil cells were performed using a standard four-probe method, with Au-wires glued to the samples using silver conductive paste. The heat capacity measurement under pressure was performed using an ac calorimetric technique, in which a heater glued onto the sample generates a small temperature oscillation $\Delta T$ by applying an ac current, and a chromel–AuFe (0.07\%) thermocouple glued to the opposite side senses an ac voltage signal proportional to $\Delta T$. The ac magnetic susceptibility measurements in a piston-cylinder-type pressure cell were carried out using an in-house-designed coil system consisting of a drive coil, a pick-up coil, and a compensation coil. The system was driven by an applied current of 0.1 mA at a frequency of 1523 Hz, and the voltage signal was detected using an SR-830 lock-in amplifier. Additional ac susceptibility measurements in a diamond anvil cell were performed up to 14 GPa using a microcoil setup. The pickup microcoil was wound from enameled Cu wire into a coil of radius of $180~\mu\mathrm{m}$, and positioned to maximize coupling to the sample in the sample chamber. A compensation coil was wound in the opposite direction outside the pickup coil to reduce background signals. The drive coil was placed on the gasket and wound using a $45~\mu\mathrm{m}$ diameter enameled wire with 250 turns. A small piece of lead (Pb), with a volume approximately one-third of the sample, was included in the pressure cell as a reference.
		The resistance, ac magnetic susceptibility, and ac heat capacity were measured using a Teslatron-PT system equipped with an Oxford $^3$He refrigerator and a Quantum Design Physical Property Measurement System (PPMS).

		\noindent \textbf{Acknowledgments} 
		We acknowledge fruitful discussions with Chao Cao, Yang Liu and Piers Coleman.
	{Work at Zhejiang University was supported by the National Key R\&D Program of China (Grant No. 2022YFA1402200, 2023YFA1406303), the National Science Foundation of China (12034017, W2511006, 12494592, 1222410, 12174332). This work has been supported by the New Cornerstone Science Foundation.}

	    \textbf{Additional information} Correspondence and requests for materials should be addressed to H. Q. Yuan (hqyuan@zju.edu.cn).

        \textbf{Author contributions} 
	H.Y. conceived the project. The crystals were grown by J.Z., Y.L, and Y.Z.. Measurements of electrical resistivity, ac susceptibility, and ac specific heat under pressure were performed by Y.Z., D.S., K.Y., and Y.H.. The experimental data were analyzed by Y.Z., D.S., Z.S., L.J., M.S., and H.Y.. Y.Z., M.S., F.S., L.J., Y.L., and H.Y. wrote the paper with input from all authors. All authors participated in discussions.

		\textbf{Competing financial interests} The authors declare no competing financial interests.

		\bibliographystyle{naturemag}
		\bibliography{ref}

@article{Wang2021,
  title = {Localized 4f-electrons in the quantum critical heavy fermion ferromagnet {CeRh$_6$Ge$_4$}},
  volume = {66},
  ISSN = {2095-9273},
  url = {http://dx.doi.org/10.1016/j.scib.2021.03.006},
  DOI = {10.1016/j.scib.2021.03.006},
  number = {14},
  journal = {Science Bulletin},
  publisher = {Elsevier BV},
  author = {Wang,  An and Du,  Feng and Zhang,  Yongjun and Graf,  David and Shen,  Bin and Chen,  Ye and Liu,  Yang and Smidman,  Michael and Cao,  Chao and Steglich,  Frank and Yuan,  Huiqiu},
  year = {2021},
  month = jul,
  pages = {1389–1394}
}

@article{Nbsn,
  title = {Specific heat of {Nb}$_3${Sn}: The case for a second energy gap},
  author = {Guritanu, V. and Goldacker, W. and Bouquet, F. and Wang, Y. and Lortz, R. and Goll, G. and Junod, A.},
  journal = {Phys. Rev. B},
  volume = {70},
  issue = {18},
  pages = {184526},
  numpages = {8},
  year = {2004},
  month = {Nov},
  publisher = {American Physical Society},
  doi = {10.1103/PhysRevB.70.184526},
  url = {https://link.aps.org/doi/10.1103/PhysRevB.70.184526}
}

@article{Askerzade2002,
  title = {On the Ginzburg-Landau analysis of the upper critical fiel d{H}c$_2$ in {MgB}$_2$},
  volume = {15},
  ISSN = {0953-2048},
  url = {http://dx.doi.org/10.1088/0953-2048/15/2/102},
  DOI = {10.1088/0953-2048/15/2/102},
  number = {2},
  journal = {Superconductor Science and Technology},
  publisher = {IOP Publishing},
  author = {Askerzade,  I N and Gencer,  A and G\"{u}\c{c}l\"{u},  N},
  year = {2002},
  month = jan,
  pages = {L13–L16}
}

@article{Pfleiderer-RMP,
  title = {Superconducting phases of $f$-electron compounds},
  author = {Pfleiderer, Christian},
  journal = {Rev. Mod. Phys.},
  volume = {81},
  issue = {4},
  pages = {1551--1624},
  numpages = {0},
  year = {2009},
  month = {Nov},
  publisher = {American Physical Society},
  doi = {10.1103/RevModPhys.81.1551},
  url = {https://link.aps.org/doi/10.1103/RevModPhys.81.1551}
}

@article{Tayama2002,
  title = {Unconventional heavy-fermion superconductor {CeCoIn}$_{5}$: dc magnetization study at temperatures down to 50 m{K}},
  author = {Tayama, T. and Harita, A. and Sakakibara, T. and Haga, Y. and Shishido, H. and Settai, R. and Onuki, Y.},
  journal = {Phys. Rev. B},
  volume = {65},
  issue = {18},
  pages = {180504},
  numpages = {4},
  year = {2002},
  month = {Apr},
  publisher = {American Physical Society},
  doi = {10.1103/PhysRevB.65.180504},
  url = {https://link.aps.org/doi/10.1103/PhysRevB.65.180504}
}

@article{Knebel2001,
  title = {Electronic properties of {CeIn}$_{3}$ under high pressure near the quantum critical point},
  author = {Knebel, G. and Braithwaite, D. and Canfield, P. C. and Lapertot, G. and Flouquet, J.},
  journal = {Phys. Rev. B},
  volume = {65},
  issue = {2},
  pages = {024425},
  numpages = {10},
  year = {2001},
  month = {Dec},
  publisher = {American Physical Society},
  doi = {10.1103/PhysRevB.65.024425},
  url = {https://link.aps.org/doi/10.1103/PhysRevB.65.024425}
}

@article{Schrder2000,
  title = {Onset of antiferromagnetism in heavy-fermion metals},
  volume = {407},
  ISSN = {1476-4687},
  url = {http://dx.doi.org/10.1038/35030039},
  DOI = {10.1038/35030039},
  number = {6802},
  journal = {Nature},
  publisher = {Springer Science and Business Media LLC},
  author = {Schr\"{o}der,  A. and Aeppli,  G. and Coldea,  R. and Adams,  M. and Stockert,  O. and L\"{o}hneysen,  H.v. and Bucher,  E. and Ramazashvili,  R. and Coleman,  P.},
  year = {2000},
  month = sep,
  pages = {351–355}
}

@article{Schuberth2016,
  title = {Emergence of superconductivity in the canonical heavy-electron metal {YbRh}$_{2}${S}i$_{2}$},
  volume = {351},
  ISSN = {1095-9203},
  url = {http://dx.doi.org/10.1126/science.aaa9733},
  DOI = {10.1126/science.aaa9733},
  number = {6272},
  journal = {Science},
  publisher = {American Association for the Advancement of Science (AAAS)},
  author = {Schuberth,  Erwin and Tippmann,  Marc and Steinke,  Lucia and Lausberg,  Stefan and Steppke,  Alexander and Brando,  Manuel and Krellner,  Cornelius and Geibel,  Christoph and Yu,  Rong and Si,  Qimiao and Steglich,  Frank},
  year = {2016},
  month = jan,
  pages = {485–488}
}

@phdthesis{Yuan2003PhD,
  author       = {H. Q. Yuan},
  title        = {Ph.D. thesis},
  school       = {Technische Universität Dresden},
  year         = {2003}
}

@article{Squire2024,
  title = {Superconductivity beyond the Conventional Pauli Limit in High-Pressure {CeSb}$_{2}$},
  author = {Squire, Oliver P. and Hodgson, Stephen A. and Chen, Jiasheng and Fedoseev, Vitaly and de Podesta, Christian K. and Weinberger, Theodore I. and Alireza, Patricia L. and Grosche, F. Malte},
  journal = {Phys. Rev. Lett.},
  volume = {131},
  issue = {2},
  pages = {026001},
  numpages = {6},
  year = {2023},
  month = {Jul},
  publisher = {American Physical Society},
  doi = {10.1103/PhysRevLett.131.026001},
  url = {https://link.aps.org/doi/10.1103/PhysRevLett.131.026001}
}

@article{rmp2012,
  title = {A common thread: The pairing interaction for unconventional superconductors},
  author = {Scalapino, D. J.},
  journal = {Rev. Mod. Phys.},
  volume = {84},
  issue = {4},
  pages = {1383--1417},
  numpages = {0},
  year = {2012},
  month = {Oct},
  publisher = {American Physical Society},
  doi = {10.1103/RevModPhys.84.1383},
  url = {https://link.aps.org/doi/10.1103/RevModPhys.84.1383}
}

@article{stewart2017,
author = {G. R. Stewart},
title = {Unconventional superconductivity},
journal = {Advances in Physics},
volume = {66},
number = {2},
pages = {75-196},
year  = {2017},
publisher = {Taylor & Francis},
doi = {10.1080/00018732.2017.1331615},

URL = {https://doi.org/10.1080/00018732.2017.1331615},
}

@article{Matsumoto2011,
title = {{Q}uantum {C}riticality {W}ithout {T}uning in the {M}ixed {V}alence {C}ompound $\beta$-{YbAlB}$_{4}$},
  volume = {331},
  ISSN = {1095-9203},
  url = {http://dx.doi.org/10.1126/science.1197531},
  DOI = {10.1126/science.1197531},
  number = {6015},
  journal = {Science},
  publisher = {American Association for the Advancement of Science (AAAS)},
  author = {Matsumoto,  Yosuke and Nakatsuji,  Satoru and Kuga,  Kentaro and Karaki,  Yoshitomo and Horie,  Naoki and Shimura,  Yasuyuki and Sakakibara,  Toshiro and Nevidomskyy,  Andriy H. and Coleman,  Piers},
  year = {2011},
  month = jan,
  pages = {316–319}
}

@article{Griveau2004,
  title = {Pressure dependence of the superconductivity in {PuCoGa}$_{5}$},
  volume = {272–276},
  ISSN = {0304-8853},
  url = {http://dx.doi.org/10.1016/j.jmmm.2003.11.056},
  DOI = {10.1016/j.jmmm.2003.11.056},
  journal = {Journal of Magnetism and Magnetic Materials},
  publisher = {Elsevier BV},
  author = {Griveau,  J.-C and Pfleiderer,  C and Boulet,  P and Rebizant,  J and Wastin,  F},
  year = {2004},
  month = may,
  pages = {154–155}
}

@article{Vargoz1998,
  title = {Upper critical field of {CeCu$_{2}$Si$_{2}$} at very high pressure},
  volume = {106},
  ISSN = {0038-1098},
  url = {http://dx.doi.org/10.1016/S0038-1098(98)00086-6},
  DOI = {10.1016/s0038-1098(98)00086-6},
  number = {9},
  journal = {Solid State Communications},
  publisher = {Elsevier BV},
  author = {Vargoz,  E. and Jaccard,  D. and Genoud,  J.Y. and Brison,  J.P. and Flouquet,  J.},
  year = {1998},
  month = jun,
  pages = {631–636}
}

@article{Bauer2011,
  title = {Localized 5f electrons in superconducting {PuCoIn}$_{5}$: consequences for superconductivity in {PuCoGa}$_{5}$},
  volume = {24},
  ISSN = {1361-648X},
  url = {http://dx.doi.org/10.1088/0953-8984/24/5/052206},
  DOI = {10.1088/0953-8984/24/5/052206},
  number = {5},
  journal = {Journal of Physics: Condensed Matter},
  publisher = {IOP Publishing},
  author = {Bauer,  E D and Altarawneh,  M M and Tobash,  P H and Gofryk,  K and Ayala-Valenzuela,  O E and Mitchell,  J N and McDonald,  R D and Mielke,  C H and Ronning,  F and Griveau,  J-C and Colineau,  E and Eloirdi,  R and Caciuffo,  R and Scott,  B L and Janka,  O and Kauzlarich,  S M and Thompson,  J D},
  year = {2011},
  month = dec,
  pages = {052206}
}

@article{valence_CCS_2004,
  title = {Signatures of valence fluctuations in {CeCu$_{2}$Si$_{2}$} under high pressure},
  author = {Holmes, Alexander T. and Jaccard, Didier and Miyake, Kazumasa},
  journal = {Phys. Rev. B},
  volume = {69},
  issue = {2},
  pages = {024508},
  numpages = {11},
  year = {2004},
  month = {Jan},
  publisher = {American Physical Society},
  doi = {10.1103/PhysRevB.69.024508},
  url = {https://link.aps.org/doi/10.1103/PhysRevB.69.024508}
}

@article{2015CeRuPO,
  title = {Avoided ferromagnetic quantum critical point in {CeRuPO}},
  author = {Lengyel, E. and Macovei, M. E. and Jesche, A. and Krellner, C. and Geibel, C. and Nicklas, M.},
  journal = {Phys. Rev. B},
  volume = {91},
  issue = {3},
  pages = {035130},
  numpages = {11},
  year = {2015},
  month = {Jan},
  publisher = {American Physical Society},
  doi = {10.1103/PhysRevB.91.035130},
  url = {https://link.aps.org/doi/10.1103/PhysRevB.91.035130}
}

@article{Aoki2001,
  title = {Coexistence of superconductivity and ferromagnetism in {URhGe}},
  volume = {413},
  ISSN = {1476-4687},
  url = {http://dx.doi.org/10.1038/35098048},
  DOI = {10.1038/35098048},
  number = {6856},
  journal = {Nature},
  publisher = {Springer Science and Business Media LLC},
  author = {Aoki,  Dai and Huxley,  Andrew and Ressouche,  Eric and Braithwaite,  Daniel and Flouquet,  Jacques and Brison,  Jean-Pascal and Lhotel,  Elsa and Paulsen,  Carley},
  year = {2001},
  month = oct,
  pages = {613–616}
}

@article{2016LaCrGe,
title = {{F}erromagnetic {Q}uantum {C}ritical {P}oint {A}voided by the {A}ppearance of {A}nother {M}agnetic {P}hase in {LaCrGe}$_{3}$ under {P}ressure},
  author = {Taufour, Valentin and Kaluarachchi, Udhara S. and Khasanov, Rustem and Nguyen, Manh Cuong and Guguchia, Zurab and Biswas, Pabitra Kumar and Bonf\`a, Pietro and De Renzi, Roberto and Lin, Xiao and Kim, Stella K. and Mun, Eun Deok and Kim, Hyunsoo and Furukawa, Yuji and Wang, Cai-Zhuang and Ho, Kai-Ming and Bud'ko, Sergey L. and Canfield, Paul C.},
  journal = {Phys. Rev. Lett.},
  volume = {117},
  issue = {3},
  pages = {037207},
  numpages = {6},
  year = {2016},
  month = {Jul},
  publisher = {American Physical Society},
  doi = {10.1103/PhysRevLett.117.037207},
  url = {https://link.aps.org/doi/10.1103/PhysRevLett.117.037207}
}

@article{Slooten2009,
title = {{E}nhancement of {S}uperconductivity near the {F}erromagnetic {Q}uantum {C}ritical {P}oint in {UCoGe}},
  author = {Slooten, E. and Naka, T. and Gasparini, A. and Huang, Y. K. and de Visser, A.},
  journal = {Phys. Rev. Lett.},
  volume = {103},
  issue = {9},
  pages = {097003},
  numpages = {4},
  year = {2009},
  month = {Aug},
  publisher = {American Physical Society},
  doi = {10.1103/PhysRevLett.103.097003},
  url = {https://link.aps.org/doi/10.1103/PhysRevLett.103.097003}
}

@article{Sidorov2003,
title = {{M}agnetic phase diagram of the ferromagnetic {K}ondo-lattice compound {CeAgSb}$_{2}$ up to 80 kbar},
  author = {Sidorov, V. A. and Bauer, E. D. and Frederick, N. A. and Jeffries, J. R. and Nakatsuji, S. and Moreno, N. O. and Thompson, J. D. and Maple, M. B. and Fisk, Z.},
  journal = {Phys. Rev. B},
  volume = {67},
  issue = {22},
  pages = {224419},
  numpages = {9},
  year = {2003},
  month = {Jun},
  publisher = {American Physical Society},
  doi = {10.1103/PhysRevB.67.224419},
  url = {https://link.aps.org/doi/10.1103/PhysRevB.67.224419}
}

@article{2008Brando,
title = {{L}ogarithmic {F}ermi-{L}iquid {B}reakdown in {NbFe}$_{2}$},
 author = {Brando, M. and Duncan, W. J. and Moroni-Klementowicz, D. and Albrecht, C. and Gr\"uner, D. and Ballou, R. and Grosche, F. M.},
  journal = {Phys. Rev. Lett.},
  volume = {101},
  issue = {2},
  pages = {026401},
  numpages = {4},
  year = {2008},
  month = {Jul},
  publisher = {American Physical Society},
  doi = {10.1103/PhysRevLett.101.026401},
  url = {https://link.aps.org/doi/10.1103/PhysRevLett.101.026401}
}

@article{Gegenwart2008,
  title = {Quantum criticality in heavy-fermion metals},
  volume = {4},
  ISSN = {1745-2481},
  url = {http://dx.doi.org/10.1038/nphys892},
  DOI = {10.1038/nphys892},
  number = {3},
  journal = {Nature Physics},
  publisher = {Springer Science and Business Media LLC},
  author = {Gegenwart,  Philipp and Si,  Qimiao and Steglich,  Frank},
  year = {2008},
  month = mar,
  pages = {186–197}
}

@article{lee2006RMP,
  title = {Doping a {{Mott}} Insulator: {{Physics}} of High-Temperature Superconductivity},
  shorttitle = {Doping a {{Mott}} Insulator},
  author = {Lee, Patrick A. and Nagaosa, Naoto and Wen, Xiao-Gang},
  date = {2006-01-06},
  journal = {Reviews of Modern Physics},
  shortjournal = {Rev. Mod. Phys.},
  volume = {78},
  number = {1},
  pages = {17--85},
  publisher = {American Physical Society},
  doi = {10.1103/RevModPhys.78.17},
  url = {https://link.aps.org/doi/10.1103/RevModPhys.78.17},
  urldate = {2024-10-15}
}

@article{Dai-rmp,
  title = {Antiferromagnetic order and spin dynamics in iron-based superconductors},
  author = {Dai, Pengcheng},
  journal = {Rev. Mod. Phys.},
  volume = {87},
  issue = {3},
  pages = {855--896},
  numpages = {42},
  year = {2015},
  month = {Aug},
  publisher = {American Physical Society},
  doi = {10.1103/RevModPhys.87.855},
  url = {https://link.aps.org/doi/10.1103/RevModPhys.87.855}
}

@Inbook{Lang2004,
author="Lang, Michael
and M{\"u}ller, Jens",
editor="Bennemann, K. H.
and Ketterson, J. B.",
title="Organic Superconductors",
bookTitle="The Physics of Superconductors: Vol. II. Superconductivity in Nanostructures, High-Tc and Novel Superconductors, Organic Superconductors",
year="2004",
publisher="Springer Berlin Heidelberg",
address="Berlin, Heidelberg",
pages="453--554",
isbn="978-3-642-18914-2",
doi="10.1007/978-3-642-18914-2_7",
url="https://doi.org/10.1007/978-3-642-18914-2_7"
}

@article{2007LohneysenRMP,
 title = {Fermi-{L}iquid {I}nstabilities at {M}agnetic {Q}uantum {P}hase {T}ransitions},
  author = {H. v. Löhneysen and A. Rosch and M. Vojta and P. Wölfle},
  year = {2007},
  journal = {Reviews of Modern Physics},
  shortjournal = {Rev. Mod. Phys.},
  volume = {79},
  number = {3},
  pages = {1015--1075},
  publisher = {American Physical Society},
  doi = {10.1103/RevModPhys.79.1015},
  url = {https://link.aps.org/doi/10.1103/RevModPhys.79.1015},
  urldate = {2024-10-15}
}

@article{park2006N,
   title = {Hidden {M}agnetism and {Q}uantum {C}riticality in the {H}eavy {F}ermion {S}uperconductor {CeRhIn}$_5$},
  author = {Park, Tuson and Ronning, F. and Yuan, H. Q. and Salamon, M. B. and Movshovich, R. and Sarrao, J. L. and Thompson, J. D.},
  year = {2006},
  journaltitle = {Nature},
  volume = {440},
  number = {7080},
  pages = {65--68},
  publisher = {Nature Publishing Group},
  issn = {1476-4687},
  doi = {10.1038/nature04571},
  url = {https://www.nature.com/articles/nature04571},
  urldate = {2024-10-15},
  langid = {english}
}

@article{shen2020N,
  title = {Strange-Metal Behaviour in a Pure Ferromagnetic {{Kondo}} Lattice},
  author = {Shen, Bin and Zhang, Yongjun and Komijani, Yashar and Nicklas, Michael and Borth, Robert and Wang, An and Chen, Ye and Nie, Zhiyong and Li, Rui and Lu, Xin and Lee, Hanoh and Smidman, Michael and Steglich, Frank and Coleman, Piers and Yuan, Huiqiu},
year = {2020},
  journaltitle = {Nature},
  volume = {579},
  number = {7797},
  pages = {51--55},
  publisher = {Nature Publishing Group},
  issn = {1476-4687},
  doi = {10.1038/s41586-020-2052-z},
  url = {https://www.nature.com/articles/s41586-020-2052-z},
  urldate = {2024-10-15},
  langid = {english}
}

@article{Saxena2000,
  title = {Superconductivity on the border of itinerant-electron ferromagnetism in {UGe}$_2$},
  author = {Saxena, S. S. and Agarwal, P. and Ahilan, K. and Grosche, F. M. and Haselwimmer, R. K. W. and Steiner, M. J. and Pugh, E. and Walker, I. R. and Julian, S. R. and Monthoux, P. and Lonzarich, G. G. and Huxley, A. and Sheikin, I. and Braithwaite, D. and Flouquet, J.},
  journal = {Nature},
  volume = {406},
  number = {6796},
  pages = {587--592},
  year = {2000},
  url = {http://dx.doi.org/10.1038/35020500},
  doi = {10.1038/35020500},
  publisher = {Springer Science and Business Media LLC}
}

@article{Luo2015,
  title = {Pressure-tuned {Q}uantum {C}riticality in the {A}ntiferromagnetic {K}ondo {S}emimetal {CeNi}$_{2-\delta}${As}$_2$},
  author = {Luo, Yongkang and Ronning, F. and Wakeham, N. and Lu, Xin and Park, Tuson and Xu, Z.-A. and Thompson, J. D.},
  journal = {Proceedings of the National Academy of Sciences},
  shortjournal = {Proc. Natl. Acad. Sci. U.S.A.},
  volume = {112},
  number = {44},
  pages = {13520--13524},
  year = {2015},
  month = {Oct},
  doi = {10.1073/pnas.1509581112},
  url = {http://dx.doi.org/10.1073/pnas.1509581112},
  issn = {1091-6490},
  publisher = {Proceedings of the National Academy of Sciences},
  langid = {english}
}

@article{su2024,
  title = {Coexistence of ferromagnetism and cluster glass behavior in {Ce$_{5}$CoGe$_{2}$}},
  author = {Su, Dajun and Zhang, Jiawen and Zhang, Yongjun and Shan, Zhaoyang and Zhang, Yanan and Smidman, Michael and Jiao, Lin and Liu, Yu and Yuan, Huiqiu},
  journal = {Phys. Rev. B},
  volume = {110},
  issue = {14},
  pages = {144432},
  numpages = {8},
  year = {2024},
  month = {Oct},
  publisher = {American Physical Society},
  doi = {10.1103/PhysRevB.110.144432},
  url = {https://link.aps.org/doi/10.1103/PhysRevB.110.144432}
}

@article{Mathur1998,
  title={Magnetically mediated superconductivity in heavy fermion compounds},
  author={Mathur, N. D. and Grosche, F. M.and Julian, S. R. and Walker, I. R. and Freye, D. M. and Haselwimmer, R. K. W. and  Lonzarich, G. G.},
  journal={Nature},
  volume={394},
  number={6688},
  pages={39--43},
  year={1998},
  publisher={Nature Research},
  doi = {10.1038/27838},
  url = {http://dx.doi.org/10.1038/27838}
}

@article{BrandoRMP2016,
  title = {Metallic quantum ferromagnets},
  author = {Brando, M. and Belitz, D. and Grosche, F. M. and Kirkpatrick, T. R.},
  journal = {Rev. Mod. Phys.},
  volume = {88},
  issue = {2},
  pages = {025006},
  numpages = {71},
  year = {2016},
  month = {May},
  publisher = {American Physical Society},
  doi = {10.1103/RevModPhys.88.025006},
  url = {https://link.aps.org/doi/10.1103/RevModPhys.88.025006}
}

@article{Eiling1981,
doi = {10.1088/0305-4608/11/3/010},
url = {https://dx.doi.org/10.1088/0305-4608/11/3/010},
year = {1981},
month = {mar},
publisher = {},
volume = {11},
number = {3},
pages = {623},
author = {A Eiling and  J S Schilling},
title = {Pressure and temperature dependence of electrical resistivity of {P}b and {S}n from 1-300~{K} and 0-10~{GPa}-use as continuous resistive pressure monitor accurate over wide temperature range; superconductivity under pressure in {Pb}, {Sn} and {In}},
journal = {Journal of Physics F: Metal Physics},

}

@article{Yuan2003,
	author = {H. Q. Yuan  and F. M. Grosche  and M. Deppe  and C. Geibel  and G. Sparn  and F. Steglich },
	title = {Observation of {Two} {Distinct} {Superconducting} {Phases} in {CeCu$_{2}$Si$_{2}$}},
	journal = {Science},
	volume = {302},
	number = {5653},
	pages = {2104-2107},
	year = {2003},
	doi = {10.1126/science.1091648},
	URL = {https://www.science.org/doi/abs/10.1126/science.1091648}
}

@article{Yuan2006,
  title = {Non-Fermi Liquid States in the Pressurized {CeCu$_2$(Si$_{1-x}$Ge$_x$)$_2$} System: Two Critical Points},
  author = {Yuan, H. Q. and Grosche, F. M. and Deppe, M. and Sparn, G. and Geibel, C. and Steglich, F.},
  journal = {Phys. Rev. Lett.},
  volume = {96},
  issue = {4},
  pages = {047008},
  numpages = {4},
  year = {2006},
  month = {Feb},
  publisher = {American Physical Society},
  doi = {10.1103/PhysRevLett.96.047008},
  url = {https://link.aps.org/doi/10.1103/PhysRevLett.96.047008}
}

@article{mao1986rubycalibration,
	author = {Mao, H. K. and Xu, J. and Bell, P. M.},
	title = {Calibration of the ruby pressure gauge to 800 kbar under quasi-hydrostatic conditions},
	journal = {Journal of Geophysical Research: Solid Earth},
	volume = {91},
	number = {B5},
	pages = {4673-4676},
	doi = {https://doi.org/10.1029/JB091iB05p04673},
	url = {https://agupubs.onlinelibrary.wiley.com/doi/abs/10.1029/JB091iB05p04673},
	year = {1986}
}

@article{Sakai2012,
  title = {{S}uperconductivity in the {F}erroquadrupolar {S}tate in the {Q}uadrupolar {K}ondo {L}attice {PrTi}$_{2}${Al}$_{20}$},
  volume = {81},
  ISSN = {1347-4073},
  url = {http://dx.doi.org/10.1143/JPSJ.81.083702},
  DOI = {10.1143/jpsj.81.083702},
  number = {8},
  journal = {Journal of the Physical Society of Japan},
  publisher = {Physical Society of Japan},
  author = {Sakai,  Akito and Kuga,  Kentaro and Nakatsuji,  Satoru},
  year = {2012},
  month = aug,
  pages = {083702}
}

@article{steppke2013,
 title = {Ferromagnetic {Q}uantum {C}ritical {P}oint in the {H}eavy-{F}ermion {M}etal {YbNi$_4$}({P}$_{1-x}${A}s$_x$)$_2$},
  author = {Steppke, Alexander and Küchler, Robert and Lausberg, Stefan and Lengyel, Edit and Steinke, Lucia and Borth, Robert and Lühmann, Thomas and Krellner, Cornelius and Nicklas, Michael and Geibel, Christoph and Steglich, Frank and Brando, Manuel},
  year = {2013},
  journal = {Science},
  volume = {339},
  number = {6122},
  pages = {933--936},
  publisher = {American Association for the Advancement of Science},
  doi = {10.1126/science.1230583},
  url = {https://www.science.org/doi/10.1126/science.1230583},
  urldate = {2024-10-15}
}

@article{Michael_RMP,
  title = {Colloquium: Unconventional fully gapped superconductivity in the heavy-fermion metal {CeCu}$_2${Si}$_2$},
  author = {Smidman, Michael and Stockert, Oliver and Nica, Emilian M. and Liu, Yang and Yuan, Huiqiu and Si, Qimiao and Steglich, Frank},
  journal = {Rev. Mod. Phys.},
  volume = {95},
  issue = {3},
  pages = {031002},
  numpages = {29},
  year = {2023},
  month = {Sep},
  publisher = {American Physical Society},
  doi = {10.1103/RevModPhys.95.031002},
  url = {https://link.aps.org/doi/10.1103/RevModPhys.95.031002}
}

@article{Tomita_YbAlB_PRESSURE,
  title = {Pressure-induced magnetic transition exceeding 30~K in the {Yb}-based heavy-fermion $\beta$-{YbAlB}$_4$},
  author = {Tomita, Takahiro and Kuga, Kentaro and Uwatoko, Yoshiya and Nakatsuji, Satoru},
  journal = {Phys. Rev. B},
  volume = {94},
  issue = {24},
  pages = {245130},
  numpages = {6},
  year = {2016},
  month = {Dec},
  publisher = {American Physical Society},
  doi = {10.1103/PhysRevB.94.245130},
  url = {https://link.aps.org/doi/10.1103/PhysRevB.94.245130}
}

@article{2001StewartRMP,
title = {Non-{Fermi}-liquid {B}ehavior in $d$- and $f$-{E}lectron {M}etals},
  author = {G. R. Stewart},
  year = {2001},
  journal = {Reviews of Modern Physics},
  shortjournal = {Rev. Mod. Phys.},
  volume = {73},
  number = {4},
  pages = {797--855},
  publisher = {American Physical Society},
  doi = {10.1103/RevModPhys.73.797},
  url = {https://link.aps.org/doi/10.1103/RevModPhys.73.797},
  urldate = {2024-10-15}
}

@article{WHH,
  title = {Temperature and {Purity} {Dependence} of the {Superconducting} {Critical} {Field}, ${H}_{c2}$. III. {Electron} {Spin} and {Spin}-{Orbit} {Effects}},
  author = {Werthamer, N. R. and Helfand, E. and Hohenberg, P. C.},
  journal = {Phys. Rev.},
  volume = {147},
  issue = {1},
  pages = {295--302},
  numpages = {0},
  year = {1966},
  month = {Jul},
  publisher = {American Physical Society},
  doi = {10.1103/PhysRev.147.295},
  url = {https://link.aps.org/doi/10.1103/PhysRev.147.295}
}

@article{Park2008,
  doi = {10.1073/pnas.0801873105},
  url = {https://doi.org/10.1073/pnas.0801873105},
  year = {2008},
  month = may,
  publisher = {Proceedings of the National Academy of Sciences},
  volume = {105},
  number = {19},
  pages = {6825--6828},
  author = {T. Park and M. J. Graf and L. Boulaevskii and J. L. Sarrao and J. D. Thompson},
  title = {Electronic duality in strongly correlated matter},
  journal = {Proceedings of the National Academy of Sciences}
}

@article{knebel2008,
  doi = {10.1143/jpsj.77.114704},
  title={The quantum critical point in {CeRhIn$_{5}$}: {A} resistivity study},
  author={Knebel, Georg and Aoki, Dai and Brison, Jean-Pascal and Flouquet, Jacques},
  journal={Journal of the Physical Society of Japan},
  volume={77},
  number={11},
  pages={114704},
  year={2008},
  publisher={The Physical Society of Japan}
}

@article{sullow1999,
  title = {Doniach {P}hase {D}iagram, {R}evisited: {F}rom {F}erromagnet to {F}ermi {L}iquid in {P}ressurized {CeRu}$_2${Ge}$_2$},
  shorttitle = {Doniach {{Phase Diagram}}, {{Revisited}}},
  author = {Süllow, S. and Aronson, M. C. and Rainford, B. D. and Haen, P.},
   year = {1999},
  journal = {Physical Review Letters},
  shortjournal = {Phys. Rev. Lett.},
  volume = {82},
  number = {14},
  pages = {2963--2966},
  issn = {0031-9007, 1079-7114},
  doi = {10.1103/PhysRevLett.82.2963},
  url = {https://link.aps.org/doi/10.1103/PhysRevLett.82.2963},
  urldate = {2024-10-15},
  langid = {english}
}

		\clearpage

\newpage
\clearpage

	\clearpage

	\end{document}